\documentclass[sigconf,screen]{acmart}

\usepackage[linesnumbered,ruled,vlined]{algorithm2e}
\usepackage{float}
\usepackage[colorinlistoftodos]{todonotes} 
\usepackage{subcaption}

\copyrightyear{2026}
\acmYear{2026}
\setcopyright{cc}
\setcctype{by}
\acmConference[LAK 2026]{LAK26: 16th International Learning Analytics and Knowledge Conference}{April 27-May 01, 2026}{Bergen, Norway}
\acmBooktitle{LAK26: 16th International Learning Analytics and Knowledge Conference (LAK 2026), April 27-May 01, 2026, Bergen, Norway}
\acmPrice{}
\acmDOI{10.1145/3785022.3785041}
\acmISBN{979-8-4007-2066-6/2026/04}

\begin{document}

\title[How Human Tutoring Interactions Shape Engagement in Online Learning]{Brief but Impactful: How Human Tutoring Interactions Shape Engagement in Online Learning}

\author{Conrad Borchers}
\orcid{0000-0003-3437-8979}
\affiliation{
\department{Human-Computer Interaction Institute, School of Computer Science}
    \institution{Carnegie Mellon University}
    \city{Pittsburgh, PA}
    \country{USA}
}
\email{cborcher@cs.cmu.edu}

\author{Ashish Gurung}
\orcid{0000-0001-7003-1476}
\affiliation{
\department{Human-Computer Interaction Institute, School of Computer Science}
    \institution{Carnegie Mellon University}
    \city{Pittsburgh, PA}
    \country{USA}
}
\email{agurung@cmu.edu}

\author{Qinyi Liu}
\orcid{0009-0003-4973-0901}
\affiliation{  
    \department{Centre for the Science of Learning \& Technology (SLATE)}
    \institution{University of Bergen}
    \city{Bergen}
    \country{Norway}
}
\email{qinyi.liu@uib.no}

\author{Danielle R. Thomas}
\orcid{0000-0001-8196-3252}
\affiliation{
\department{Human-Computer Interaction Institute, School of Computer Science}
  \institution{Carnegie Mellon University}
  \city{Pittsburgh, PA}
  \country{USA}
}
\email{drthomas@cmu.edu}

\author{Mohammad Khalil}
\orcid{0000-0002-6860-4404}
\affiliation{
    \department{Centre for the Science of Learning \& Technology (SLATE)}
    \institution{University of Bergen}
    \city{Bergen}
    \country{Norway}
}
\email{mohammad.khalil@uib.no}

\author{Kenneth R. Koedinger}
\orcid{0000-0002-5850-4768}
\affiliation{
\department{Human-Computer Interaction Institute, School of Computer Science}
  \institution{Carnegie Mellon University}
  \city{Pittsburgh, PA}
  \country{USA}
}
\email{koedinger@cmu.edu}

\renewcommand{\shortauthors}{Borchers et al.}

\newcommand{\revision}[1]{\textcolor{black}{#1}}

\begin{abstract}
Learning analytics can guide human tutors to efficiently address motivational barriers to learning that AI systems struggle to support. Students become more engaged when they receive human attention. However, what occurs during short interventions, and when are they most effective? We align student–tutor dialogue transcripts with \emph{MATHia} tutoring system log data to study brief human-tutor interactions on Zoom drawn from 2,075 hours of 191 middle school students' classroom math practice. Mixed-effect models reveal that engagement, measured as successful solution steps per minute, is higher during a human-tutor visit and remains elevated afterward. Visit length exhibits diminishing returns: engagement rises during and shortly after visits, irrespective of visit length. Timing also matters: later visits yield larger immediate lifts than earlier ones, though an early visit remains important to counteract engagement decline. We create analytics that identify which tutor-student dialogues raise engagement the most. Qualitative analysis reveals that interactions with concrete, stepwise scaffolding with explicit work organization elevate engagement most strongly. We discuss implications for resource-constrained tutoring, prioritizing several brief, well-timed check-ins by a human tutor while ensuring at least one early contact. Our analytics can guide the prioritization of students for support and surface effective tutor moves in real-time.
\end{abstract}

\ccsdesc[500]{Computing methodologies~Modeling and simulation}
\ccsdesc[300]{Applied computing~Education}

\keywords{Engagement; Hybrid human–AI tutoring; Intelligent tutoring systems; Dialogue analysis; K-12}
 
\maketitle

\section{Introduction}

AI-based tutoring systems can provide effective, adaptive practice at scale by guiding students step-by-step, delivering immediate feedback, and offering targeted hints \cite{ritter2016mastery,koedinger2023astonishing}. However, they remain limited in addressing the motivational and emotional factors that shape engagement \cite{baker2016stupid}. In hybrid classrooms, human tutors complement AI tutors by diagnosing affective states, restoring attention, and offering encouragement or clarification when needed \cite{holstein2018student,thomas2024improving,wang2024tutor,gurung2025human}. This practice often takes the form of \emph{hybrid human–AI tutoring}, in which students primarily learn through the AI system while tutors periodically “visit” for brief, targeted conversations. Such visits address both content and motivational needs, from scaffolding steps to sustaining persistence \cite{thomas2024improving,gurung2025human}. A central challenge, however, is how to allocate scarce tutoring time: \emph{when} to intervene, \emph{how long} to stay with a student, and \emph{which dialogue moves} best foster engagement.

Analytics that guide short tutor interventions can temporarily raise student engagement in classroom sessions \cite{holstein2018student,jin2025who,karumbaiah2023spatiotemporal}. Yet learning analytics lacks fine-grained evidence on the \emph{dynamics} of such boosts, including their relation to visit \emph{duration} and \emph{timing}. It also remains unclear which dialogue features between tutors and students yield stronger gains, and how these patterns interact with the broader decline in engagement typically seen among students who are visited less frequently. Addressing these gaps requires fine-grained analyses of \emph{traces} and system \emph{logs} that align tutor interactions with moment-to-moment measures of student activity to strengthen the impact of tutoring on learning.

To address these gaps, we align tutor–student dialogue with tutoring-system log data in middle school math. Using MATHia minute-level traces time-matched to Zoom tutor transcripts, we analyze 2,075 hours of hybrid tutoring from 191 students. Our contribution is twofold: first, a reproducible method to label tutor \emph{visits} by matching dialogue onsets with system events within a one-minute window, allowing within-session modeling of engagement while retaining full activity timelines; second, a mixed-methods approach linking model-estimated engagement gains to dialogue content through sentence embeddings and qualitative coding. Combining trace data with dialogue analysis, we advance methods for examining \emph{how, when, and why} tutor interventions in hybrid classrooms yield stronger engagement boosts.

This work has practical and theoretical importance. In practice, the dosage of human tutoring has been shown to improve student outcomes \cite{nickow2020impressive,kraft2021blueprint}; understanding how tutors maximize impact in short visits informs attention-allocation in resource-limited classrooms. Theoretically, our findings clarify how human support counters within-session declines and reveal dialogue features that heighten engagement. For learning analytics, we show how fine-grained alignment of multimodal data can generate actionable insights for real-time support, such as teacher dashboards \cite{wang2024tutor,holstein2018student}. We ask:

\begin{itemize}
    \item \textbf{RQ1.} What is the impact of human tutor intervention on student engagement during and after visits?
    \item \textbf{RQ2.} Does the \emph{duration} of visits moderate their effects on engagement?
    \item \textbf{RQ3.} Does the \emph{timing} of visits within a session moderate their effects (controlling for duration)?
    \item \textbf{RQ4.} Which dialogue characteristics are associated with stronger engagement uplift?
\end{itemize}

\section{Background and Motivation}

A central construct in hybrid human--AI tutoring is \emph{engagement}, recognized as essential for sustaining attention and effort that drive learning. Learning analytics captures it through fine-grained activity traces. Before turning to attention allocation and dialogue, we examine how engagement has been defined in computer-based practice and why it offers a crucial lens for assessing tutor interventions guided by learning analytics.

\subsection{Engagement in Computer-Based Practice}

Engagement is a prerequisite for effective learning in computer-based environments, encompassing behavioral, emotional, and cognitive dimensions. Disengagement may appear as reduced effort, random guessing, or off-task behavior \cite{baker2010better,baker2006adapting}. Such states are not always rooted in misunderstanding: students capable of solving a problem may disengage due to motivational or affective barriers. Yet disengagement consistently predicts lower outcomes \cite{rodrigo2008use,pardos2013affective} and undermines tutoring systems that rely on active participation to deliver personalized support.

Within learning analytics, engagement has become a central concern because it signals when instructional interventions are needed: detecting disengagement early enables support before learning opportunities are lost \cite{clow2012lac,siemens2013la}. A large body of work has developed models that infer disengagement from traces such as gaming the system \cite{baker2006adapting}, affective states like boredom or frustration \cite{pardos2013affective,miller2014boredom}, and indicators of productive struggle \cite{young2024productive}. Increasingly, these analytics are being integrated into dashboards and orchestration tools that help teachers and tutors intervene effectively in real time \cite{holstein2018student}.

In intelligent tutoring systems, engagement is often measured through the completion of learning steps with accuracy, a robust proxy for both motivation and progress \cite{koedinger2023astonishing,borchers2024combining}. Each step represents an opportunity to learn from feedback and thus reflects how much content a student masters over time \cite{koedinger2023astonishing}. Recent studies show that interventions aimed at sustaining engagement—such as goal setting or timely tutor support—can enhance learning outcomes by increasing students’ time and effort in practice \cite{borchers2025engagement}. In hybrid human–AI settings, pairing step-completion measures with dialogue analysis captures how tutors influence moment-to-moment engagement dynamics.

The present study addresses this gap by treating engagement not only as a predicted outcome but as a dynamic process unfolding minute by minute in hybrid tutoring\revision{, where several tutors attend to students who learn with technology in class through video conferencing. This model addresses the reality that classroom teachers cannot simultaneously provide individualized attention to every student while informing analytics-based interventions that could guide scarce human attention in technology-supported classrooms.} We align step-level measures with tutor dialogue to capture both immediate and sustained effects of visits, examine how their length and timing shape these effects, and identify dialogue features that distinguish higher- from lower-impact episodes. In doing so, we extend engagement detection toward explanatory accounts that reveal how human support, potentially enhanced by learning analytics, can counter within-session declines and sustain productive practice in AI-driven environments.

\subsection{Analytics of Attention-Allocation in Classrooms}

Research on classroom attention has a long history. Scholars have studied how students distribute attention during lessons \cite{fisher2014visual}, while parallel work has examined how teachers allocate attention across students, with implications for engagement, learning, and professional development \cite{Bozkir2025}. With the rise of digital learning and trace data, attention allocation has become a central focus in learning analytics \cite{Yan2021,karumbaiah2023spatiotemporal,jin2025who}. By analyzing behavioral and emotional data, analytics can offer teachers insights to optimize their attention and enhance classroom learning \cite{Godwin2022}.

During classroom practice, engagement often fluctuates and declines \cite{baker2006adapting,gurung2025starting}. Attention analytics can help counteract this by detecting early signs of disengagement, identifying students in need of support, and enabling timely interventions \cite{Godwin2022}. Wearable tools such as Lumilo let teachers monitor students’ real-time states and intervene proactively \cite{holstein2018student,karumbaiah2023spatiotemporal,Karumbaiah2023TeacherNoticing}. Likewise, blended classrooms studies show that well-timed teacher visits can restore focus and boost engagement \cite{Jeffrey2014BlendedLearning}.

This study advances prior work by analyzing a unique data set that combines log-based engagement measures with co-occurring tutor–student dialogue. By aligning minute-level engagement traces with the timing and duration of tutor visits, we examine how attention evolves during sessions and which scheduling choices are most effective. Moving beyond monitoring, we provide evidence of how human attention counters engagement decline and offer heuristics for when and how tutors should allocate limited time in resource-constrained classrooms.

\subsection{Dialogue and Tutoring Analytics}

Human attention and interaction are known to enhance student engagement. Classroom studies showed that teacher presence increases on-task behavior \cite{person2003simulating}, while spatiotemporal learning analytics revealed how teacher movement and noticing create learning opportunities \cite{karumbaiah2023spatiotemporal}. Dialogue research has further highlighted how instructional moves such as scaffolding, questioning, and praise support learning \cite{vail2014identifying,chi2001learning}. Extending this work, multimodal learning analytics integrates traces of student learning with dialogue analysis to capture how social and cognitive processes unfold \cite{d2015multimodal,blikstein2014multimodal}.

Recent advances in dialogue act modeling enable fine-grained analysis of tutoring interactions at scale. Tutor CoPilot, for example, integrates dialogue act detection with real-time analytics to guide tutors in allocating attention and responding effectively to learners \cite{wang2024tutor}. Dialogue act modeling has also been combined with skill modeling to identify conversational exchanges in AI-supported peer tutoring that are most strongly associated with learning efficiency, offering an early instance of linking dialogue acts to system log data \cite{borchers2024combining}. These studies show how dialogue analytics connect log interaction patterns with broader outcomes, providing explanatory insights and real-time support for orchestration.

We advance this line of work by linking tutor–student dialogue to measurable engagement uplift in an AI tutoring environment. We align transcripts with moment-to-moment system logs to identify conversational moves that coincide with student productivity gains. This integration bridges dialogue analytics and engagement modeling, creating new insights into what forms of scaffolding, questioning, and work organization most effectively sustain engagement.

\section{Methods}

\subsection{Study Context} 

Students participated in a hybrid human–AI tutoring program in which human tutors and an AI system worked collaboratively to provide individualized academic support. Human tutors focused on motivational, relational, and math content-related guidance, while the AI system provided adaptive instruction through Carnegie Learning's MATHia platform (formerly Cognitive Tutor \cite{ritter2007cognitive}), designed for grades 6–12. During school hours, students used MATHia twice a week for about 45-minute sessions, and remote tutors joined via Zoom to provide real-time support. Tutors were trained in strategies such as responding to errors, offering meaningful praise, and scaffolding problem solving to prepare them for both instructional and motivational guidance in a virtual setting. Tutors were given guidelines on when and how to intervene and instructed to prioritize students who were idle, with low time on task, and showed limited progress in the system. At the same time, students were able to virtually raise their hands to be visited and supported by tutors.

The program was implemented in the fall semester of the 2024–2025 school year with N = 191 7th (83) and 8th-grade (108) students from a mid-Atlantic U.S. middle school. Tutoring was delivered by N = 17 trained undergraduate and graduate tutors from a U.S. university. This study was approved by the university’s Institutional Review Board (IRB). Informed consent and student assent were obtained prior to participation. All transcripts and log data were de-identified for data analysis, with anonymized IDs replacing personal information to ensure confidentiality, and secure data storage adhered to safety and security protocols. Schools were recruited through direct outreach, and participating sites implemented the program during the school day. For transparency and replication, the data analysis code is provided in this study’s supplementary Git repository.\footnote{\url{https://github.com/conradborchers/brief-but-impactful-lak26}}

\subsection{Dataset}
The dataset includes three primary data sources: MATHia transaction logs, Zoom session logs, and video transcripts of student–tutor interactions. The Zoom logs and video recordings capture tutor visits to individual breakout rooms, where tutors check in on student progress and provide support. These recordings were transcribed using WhisperX~\cite{bain2023whisperx} to enable deeper analysis of human tutoring through dialogue transcripts. Together, these data sources were integrated to examine student behavior and learning in relation to both AI-driven practice and human-delivered support. 

The MATHia transaction logs include detailed information on each student’s activity, including student ID, lesson ID, problem ID, step ID, and timestamped records of individual actions such as problem attempts and help requests. Each action also contains relevant metadata, including correctness, error information, and the type of support accessed, enabling fine-grained analysis of student learning behaviors and progress over time. The fall usage resulted in 1,022,884 transactions from the 191 students in grades 7 and 8.

The Zoom session logs include metadata for both the main room and individual breakout rooms, with precise timestamps indicating when each student and tutor entered and exited a breakout room. These logs enable us to reconstruct the timing and duration of tutor-student interactions and align them with MATHia activity (using timestamps), allowing for analysis of student behavior both during and outside of tutor presence. As students are allowed to use their own Zoom alias, some personalized their names in ways that we could not match (e.g., ``why did the chicken cross the road''), whereas others had similar names (e.g., two students with the same first name), and as such, several students could not be matched. Among the 191 students, our human coders were able to reliably match 140 students to log data (73.3\% match rate).

Finally, the video transcripts provide detailed records of tutor-student interactions during breakout room visits, including the content of each utterance, speaker diarization (i.e., who spoke), and timing relative to the start of the video. Each utterance is annotated with its start time and duration relative to the video length. Since Zoom only records video while a tutor is inside a breakout room—not while they are in the main room—Zoom session logs are used to determine when tutors entered and exited each room, while transcripts provide insights into the content of interactions during those segments. For example, if a tutor spends 20 minutes rotating through breakout rooms and an additional 5 minutes in the main room conferring with another tutor or the teacher, only the 20 minutes spent in breakout rooms would appear in the video transcript. In contrast, a fellow tutor who remains in breakout rooms for the entire 25-minute session would have a complete 25-minute transcript.

\begin{figure*}[!ht]
\centering
  \includegraphics[width=.94\linewidth]{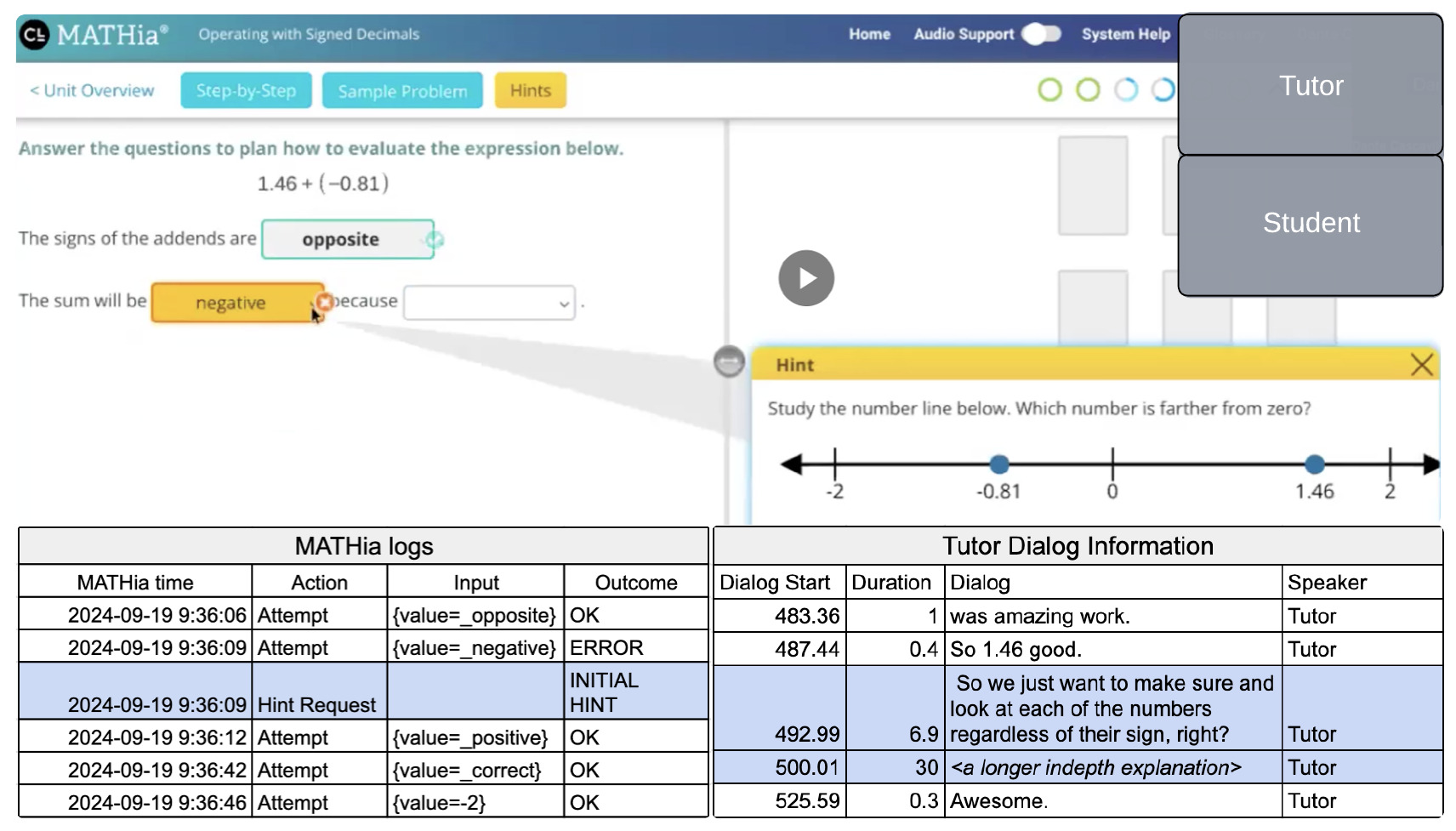}
  \caption{Visual representation of a student receiving remote tutor support while using MATHia. The Zoom screen capture (top half) is from a tutor visit, with the tutor and student (top right) grayed out to maintain anonymity. The bottom left shows the corresponding MATHia transaction logs, with the current log entry highlighted in blue. The bottom right presents the aligned WhisperX transcript, highlighting the tutor's instructional dialogue (also in blue) that corresponds to the student's actions.}
  \label{fig:mapped}
\end{figure*}

A visual representation of how these data were connected across sources using real student data is presented in Figure~\ref{fig:mapped}. The figure illustrates how MATHia activity logs, Zoom session metadata, and video transcripts were synchronized to capture a unified view of student learning and tutor support. This integration allows us to trace a specific student action (e.g., an incorrect attempt) and align it with the corresponding tutor intervention, as captured in the transcript and log timestamps.

\subsection{Matching Tutor Dialogue to In-System Activity}
\label{sec:methods:data-alignment}

We aligned tutor--student dialogue with students' activity in the tutoring system to identify when a \emph{tutor dialogue intervention} occurred during ongoing work. For each student, we chronologically ordered (a) dialogue starts and (b) system events that reflect student action (e.g., problem/step attempts and outcomes). For every dialogue start, we searched \emph{forward} for the nearest subsequent system event by the same student occurring within a one-minute window. If such an event existed, the dialogue was attached to that event; when multiple dialogue turns fell within the window for the same event, we merged their contents in temporal order and treated them as one intervention linked to that event.

We define a \textbf{tutor dialogue intervention (visit)} in the log data as an event-level label assigned to any system event for which at least one tutor--student dialogue turn began within the preceding one minute for the same student. Multiple turns mapped to the same event, as well as multiple log data events with dialogue in a row, constitute a single intervention. Importantly, we retained the \emph{entire} system log timeline for analysis. Events with at least one matched dialogue turn were labeled \emph{visit} (tutor dialogue intervention). Events without any matched turn remained in the dataset and were labeled \emph{no visit}. This allows us to analyze both activity during collaboration/visits and activity in episodes without tutor intervention, as well as the effect of visit duration and timing on student engagement. We defined the \textbf{duration of an intervention} as the temporal span of the matched dialogue turns associated with a single event, i.e., the difference between the timestamps of the last and first matched dialogue turns of a visit.

This procedure yielded a dialogue match rate of 73.4\% to log data, with about 25\% of tutor dialogue interventions happening without any student actions in the tutoring system within 1 minute of a dialogue start. This happens when tutors, for example, only check in with the student to greet them or give them a short lecture to briefly interact on a math topic. For the purposes of this study, we excluded such visits, as we were interested in measuring the immediate impact of visits on students' math practice. However, we also confirmed the robustness of our results with a more lenient match window of 5 minutes (84.5\% coverage).

\revision{We treat \textbf{completed problem-solving steps} as our key engagement outcome of interest in the present study. In MATHia, problems comprise several solution steps, to which students have to construct correct solution attempts to advance to the next problem \cite{ritter2013predicting}. MATHia tutors students on each of these steps, providing immediate correctness feedback and as-needed instruction through hints and error-specific feedback messages. Depending on the math problem, steps can refer to individual transformations or operations, such as subtracting an integer from both sides in a multi-step equation-solving problem. Successful step counts are an established engagement metric in intelligent tutoring systems research. When students spend a similar amount of time using the tutoring system in class, step completion counts provide a more accurate indicator of active, on-task engagement and better predictor of learning than time-based measures (e.g., \cite{ritter2013predicting,koedinger2023astonishing}). Importantly, in tutoring systems like MATHia, where students receive hints and feedback, every step is eventually solved correctly. Therefore, this engagement measure includes (not excludes) errors and the productive use of feedback that leads to success.}

\subsection{Data Analysis Methods}

\subsubsection{RQ1: What is the Impact of Human Tutor Visits on Engagement?}

To estimate the within-session impact of brief tutor visits on student engagement (RQ1), we modeled minute-level step completions while students worked in the tutoring system. We retained the \emph{entire} event log timeline for analysis (both with and without tutor intervention), and identified tutor dialogue interventions (“visits”) by aligning tutor--student dialogue starts to system events within a one-minute forward window (see Section~\ref{sec:methods:data-alignment}). Moments with at least one matched dialogue turn were labeled \emph{visit}; all other moments were labeled \emph{no visit}.

We fit a generalized linear mixed-effects model with a Poisson likelihood and log link, using a random intercept for each student to account for repeated measurements and individual baseline engagement \cite{bolker2015linear}. The outcome \(Y_{it}\) is the count of successfully completed steps for student \(i\) at minute \(t\).

\begin{equation}
\begin{aligned}
\log Y_{it}
&= \beta_0
+ \beta_1\, \mathrm{MinsSinceStart}_{it} \\
&\quad + \beta_2\, \mathrm{DuringTutoring}_{it} \\
&\quad + \beta_3\, \mathrm{MinsSinceLastTutoring}_{it}
+ u_i
\end{aligned}
\label{eq:rq1-model}
\end{equation}

where \(u_i\) is a student-specific random intercept, assumed to be drawn from a normal distribution with mean 0 and variance \(\sigma_u^2\). This term captures stable differences between students (e.g., baseline engagement levels), so that each student’s outcomes are modeled relative to their own starting point rather than only to the group average. The $\beta_1$ term adjusts for the general within-session time trend (e.g., warm-up or fatigue). The binary indicator $\beta_2$ captures the \emph{immediate} level change and shift in engagement during a visit relative to comparable moments without a visit. The term $\beta_3$ captures how engagement changes as time elapses after the most recent visit (i.e., post-visit persistence or decay). The reason for modeling $\beta_3$ is that visits could interrupt or change the time trend before visits occur: for instance, a negative $\beta_1$ could indicate a downward trend in engagement during classroom time, but a positive $\beta_3$ would indicate that visits counteract that trend after they have occurred.

We report fixed effects as incident-rate ratios (\textit{IRR} \(= e^\beta\)). For example, \(\mathrm{\textit{IRR}}_{\text{visit}} = e^{\beta_2}\) quantifies the multiplicative change in step-completion rate \emph{during} a visit (holding other terms constant). Models were estimated using \texttt{lme4} in R \cite{bates2015fitting}.

\subsubsection{RQ2: Does the Duration of Tutor Visits Matter?}
\label{sec:rq2-duration}

We extend Eq.~\ref{eq:rq1-model} by adding (a) the duration in minutes of the most recent tutor visit and (b) two interactions that test whether duration modifies the immediate lift during a visit and the post-visit trajectory. As in RQ1, we fit a Poisson GLMM with a log link and a student random intercept.

\begin{equation}
\begin{aligned}
\log \mu_{it} 
&= \beta_0
+ \beta_1\,\mathrm{MinsSinceStart}_{it}
+ \beta_2\,\mathrm{DuringTutoring}_{it} \\
&\quad + \beta_3\,\mathrm{MinsSinceLastTutoring}_{it}
+ \beta_4\,\mathrm{Dur}_{it} \\
&\quad + \beta_5\,\big(\mathrm{DuringTutoring}_{it} \times \mathrm{Dur}_{it}\big) \\
&\quad + \beta_6\,\big(\mathrm{MinsSinceLastTutoring}_{it} \times \mathrm{Dur}_{it}\big)
+ u_i
\end{aligned}
\end{equation}

Here, $\mathrm{Dur}_{it}$ denotes the duration (minutes) of the \emph{most recent} visit, carried forward until the next visit (and set to 0 before any visit in a session). $\beta_4$ captures the main effect of longer (vs.\ shorter) visits; $\beta_5$ tests whether the \emph{during-visit} lift grows with visit length; $\beta_6$ tests whether visit length alters the \emph{post-visit} change per minute since the visit. Effects are reported as incident-rate ratios (\textit{IRR} $=e^{\beta}$), consistent with RQ1.

\subsubsection{RQ3: Does the Timing of Tutor Visits Matter?}
\label{sec:rq3-timing}

To test whether the \emph{timing} of a visit within a session modifies its effects, we extend the RQ2 specification by adding interactions with minutes since session start (assuming that class sessions are all about 45 minutes or longer and higher values indicate visits later in the class session). We retain the RQ2 duration main effect and interactions ($\beta_4$–$\beta_6$) to adjust for heterogeneity in visit length while isolating timing effects. The reason for adding timing to a model of length is that timing is likely negatively correlated with length (i.e., visits at the end of sessions are bound by the end of the classroom period). Therefore, it is important to adjust for visit length when estimating the effects of visit timing on engagement. As in prior models, we fit a Poisson GLMM with a log link and a student random intercept.

\begin{equation}
\begin{aligned}[c]
\log \mu_{it}
&= \beta_0
+ \beta_1\,\mathrm{MinsSinceStart}_{it}
+ \beta_2\,\mathrm{DuringTutoring}_{it} \\
&\quad + \beta_3\,\mathrm{MinsSinceLastTutoring}_{it}
+ \beta_4\,\mathrm{Dur}_{it} \\
&\quad + \beta_5\,\big(\mathrm{DuringTutoring}_{it}\!\times\!\mathrm{Dur}_{it}\big) \\
&\quad + \beta_6\,\big(\mathrm{MinsSinceLastTutoring}_{it}\!\times\!\mathrm{Dur}_{it}\big) \\
&\quad + \beta_7\,\big(\mathrm{MinsSinceStart}_{it}\!\times\!\mathrm{DuringTutoring}_{it}\big) \\
&\quad + \beta_8\,\big(\mathrm{MinsSinceStart}_{it}\!\times\!
\mathrm{MinsSinceLastTutoring}_{it}\big)
+ u_i
\end{aligned}
\end{equation}

with $u_i\sim\mathcal{N}(0,\sigma_u^2)$. Here, $\mathrm{Dur}_{it}$ is the duration (minutes) of the most recent visit (carried forward until the next visit; 0 before any visit).

$\beta_7$ tests whether the \emph{immediate} during-visit lift depends on when in the session the visit occurs (later vs.\ earlier). 
$\beta_8$ tests whether the \emph{post-visit} trajectory per minute since the last visit varies with when the visit occurs. 
As in RQ1–RQ2, we report incident-rate ratios (\textit{IRR} $=e^\beta$).

For all statistical models and tests (RQ1–RQ3), we verified distributional and modeling assumptions using standard diagnostics: Q–Q plots for residual normality, residuals vs. fitted and scale–location plots for linearity and homoscedasticity, and residuals vs. leverage plots (with Cook’s distance) for influential cases.

\subsubsection{RQ4: Identifying Episodes with Stronger Effects}
\label{sec:rq4-randomization}

To surface tutor–student episodes whose \emph{during-visit} impact on engagement was unusually large, we constructed an episode-level dataset and computed both empirical and model-based uplift scores, then flagged the top half for qualitative analysis. We looked at during-visit effects specifically because post-visit trends showed small or negligible effect sizes (see Section \ref{sec:results}).

Using the visit definition from Section~\ref{sec:methods:data-alignment}, we aggregated contiguous minutes of tutoring into episodes per student-day. For each episode, we recorded the start time (minutes since session start), duration (in minutes), mean steps/min during the episode, number of tutor turns, and the concatenated dialogue text (all turns, temporal order, concatenated by a custom separation token). 

\paragraph{Empirical uplift.}
For each student-day, we computed a session-specific baseline (mean steps completed in the tutoring system per minute). We compared that to the students' engagement measured in steps per minute during visits. An episode’s empirical uplift was computed as a percentage and as:
\[
\mathrm{Uplift}_{\text{emp}} = 100\times\frac{\overline{\text{steps}}_{\text{visit}} - \overline{\text{steps}}_{\text{no-visit, same day}}}{\overline{\text{steps}}_{\text{no-visit, same day}}}
\]

Episodes were then marked as having a strong \emph{strong effect} if their associated empirical uplift was in the top 50\% of uplifts (median split). With this separation, we then analyzed student-tutor dialogue transcriptions for high vs. low uplift scores in student engagement.

\paragraph{Randomization test of dialogue differences}

We tested whether transcripts from \emph{high-uplift} episodes are distinguishable from those with \textit{lower uplift} using sentence-BERT embeddings and label randomization tests, a common approach to assessing dialogue adaptivity \cite{borchers2025can}. Each episode’s concatenated dialogue was embedded, and we compared high- versus non-high episodes by computing the cosine distance between their centroid vectors. Under the null hypothesis that dialogue does not differ by uplift, labels are exchangeable; we therefore shuffled labels 5{,}000 times to build a null distribution and calculated the $p$-value as the proportion of shuffled distances that met or exceeded the observed distance. A length-stratified version shuffled only within word-count quartiles (0-25\%, 26-50\%, and so forth) to rule out trivial length effects.

As a complementary analysis, we tested whether embeddings could classify uplift status above chance. A logistic regression classifier was trained on the embeddings with five-fold student-level cross-validation and summarized by AUC, using default model hyperparameters in the \texttt{sklearn} Python package. To evaluate significance, we repeated the randomization procedure described above with permuted labels to form a null distribution of AUCs.

\paragraph{Sampling exemplar dialogues for qualitative coding}
To ground the quantitative results in concrete interaction patterns, we sampled representative episodes from highly and lowly uplifting dialogue. After embedding each episode’s concatenated dialogue with sentence-BERT (unit-normalized), we computed the mean embedding (centroid) for the \emph{high-uplift} group and for the \emph{other} group. The vector difference between these centroids defines a single \emph{separating direction} in embedding space. We then assigned each episode a score equal to its projection on this direction (dot product); intuitively, higher scores indicate dialogues that are more “high-uplift–like,” while lower scores indicate dialogues closer to the “other” centroid. We selected the top $k=15$ dialogues (highest scores) as \emph{high-uplift exemplars} and the bottom $k=15$ dialogues (lowest scores) as \emph{low-uplift exemplars}. Two researchers of the study team then collaboratively curated themes and insights from this set of examples in a thematic analysis with open coding.

\section{Results}
\label{sec:results}

\begin{figure*}[htbp]
    \centering
    \begin{subfigure}{0.48\textwidth}
        \centering
        \includegraphics[width=\textwidth]{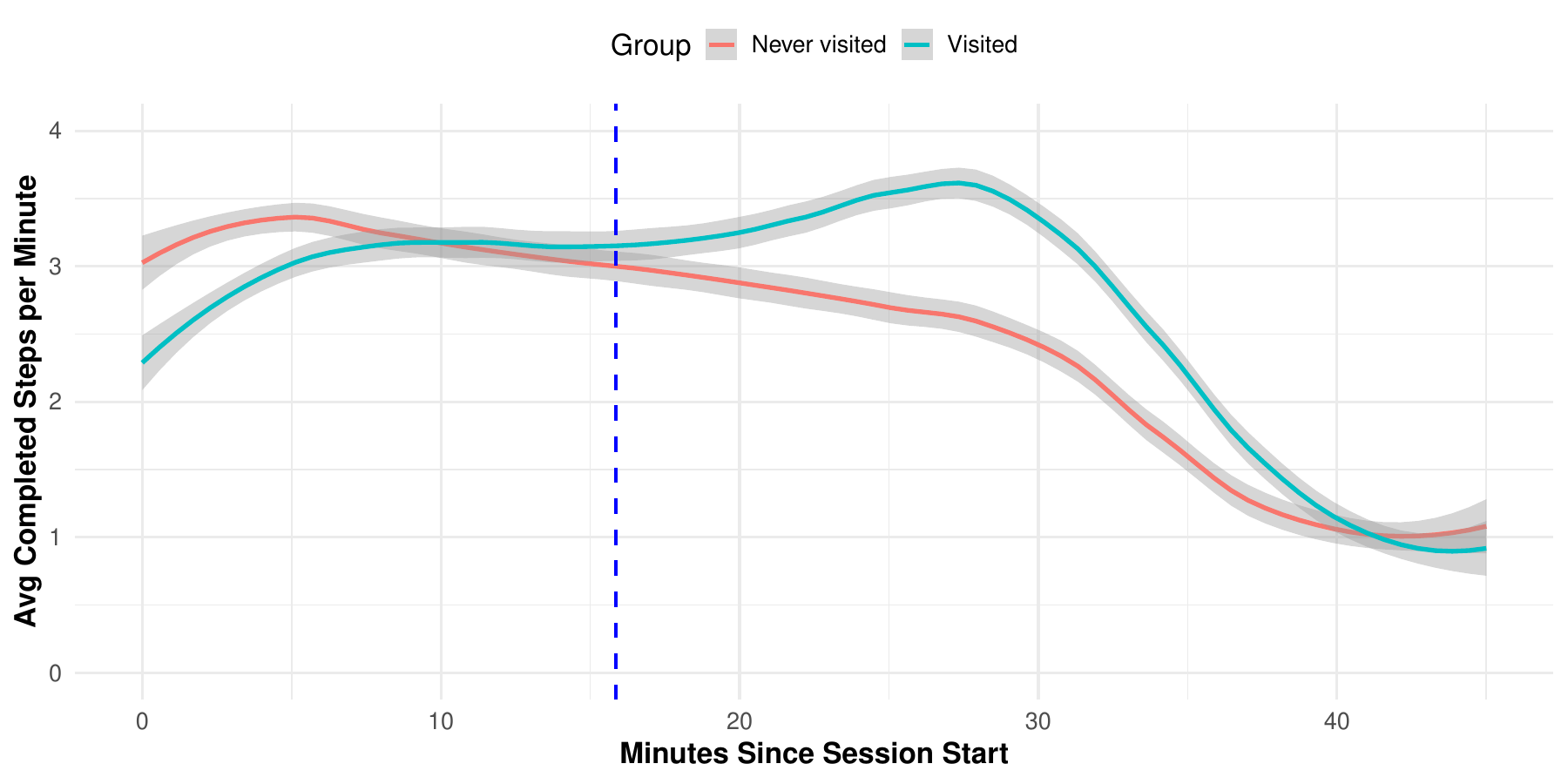}
        \caption{Successful step completion over time. The dashed line indicates the average first tutoring time for students who were visited.}
        \label{fig:steps-over-time}
    \end{subfigure}
    \hfill
    \begin{subfigure}{0.48\textwidth}
        \centering
        \includegraphics[width=\textwidth]{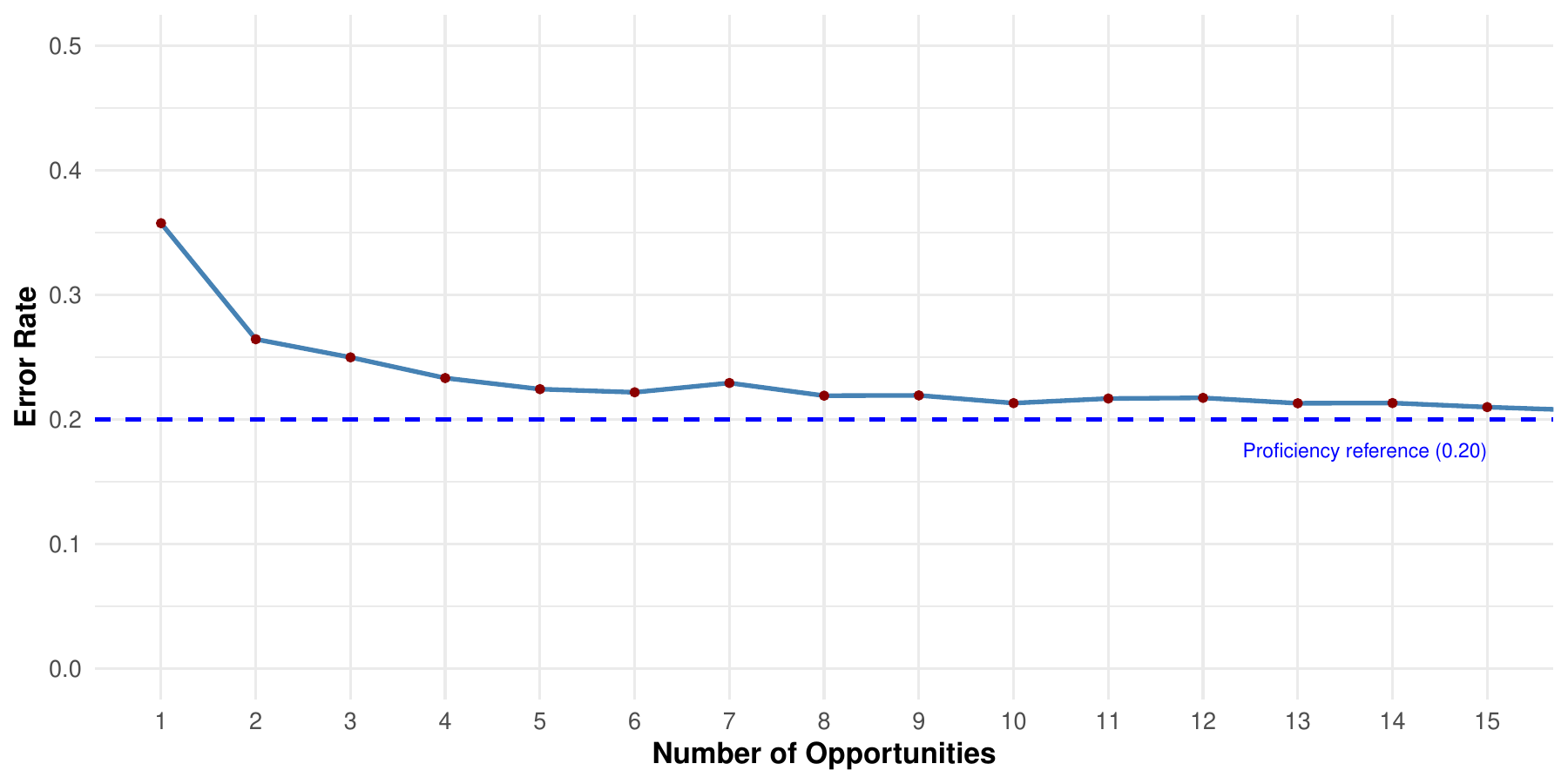}
\caption{Learning curve showing average mastery upon repeated skill practice. Flattening at higher opportunities may reflect missing data during mastery learning, which can underestimate mastery.}
        \label{fig:learning-curve}
    \end{subfigure}
    \caption{Descriptive plots validating key assumptions of our analysis. Left: Visits are associated with higher engagement after their event. Right: Students exhibited increased mastery with more learning opportunities (i.e., completed problem-solving steps), validating our choice of engagement measure.}
    \label{fig:dual-plots}
\end{figure*}

Motivating our analysis, we first replicated findings from past research showing that brief interaction periods between human tutors and students during tutoring system work temporarily increase students' engagement \cite{karumbaiah2023spatiotemporal,jin2025who}.
Descriptively, Figure~\ref{fig:steps-over-time} shows the average number of successfully completed steps per minute as a function of session time, separately for students who were never visited by a tutor and those who were visited. The solid curves are smoothed trends with confidence bands, while the dashed vertical line indicates the mean time of first tutoring among visited students. Students in both groups initially increase their rate of step completion. However, performance diverges thereafter: those who receive a visit tend to sustain or improve their rate relative to never-visited peers, particularly around the average time of intervention.

In addition to confirming the engagement effect of tutor visits, we examined how students’ error rates declined with repeated opportunities (i.e., completed problem steps) to practice the same skill (Figure~\ref{fig:learning-curve}). This analysis validates our use of completed steps as an outcome measure for two reasons: first, students typically require about seven opportunities on a given skill to reach mastery \cite{koedinger2023astonishing}; second, adaptive tutoring systems allocate more skills to students as they complete more steps. Following common practice in educational data mining \cite{rivers2016learning}, we inspected the learning curve of average error rate by opportunity and observed that students approached the widely accepted proficiency threshold of 80\% accuracy (i.e., an error rate of 0.20) \cite{koedinger2023astonishing} to more steps of the same skill they completed. This pattern confirms that student engagement, as operationalized by successful step completion, reflects meaningful progress in skill development and learning in our sample.

\subsection{Effect of Tutoring Interventions on Engagement (RQ1)}

We estimated the Poisson GLMM in Eq.~\ref{eq:rq1-model} on 123{,}439 minute-level observations from 191 students.

\begin{table}[htpb]
\centering
\caption{Poisson GLMM predicting minute-level completed steps (incidence-rate ratios, \textit{IRR}).}
\label{tab:rq1_glmm}
\begin{tabular}{lccc}
\toprule
\textbf{Predictor} & \textbf{\textit{IRR}} & \textbf{95\% \textit{CI}} & \textbf{$p$} \\
\midrule
(Intercept) & 3.32 & [3.12, 3.52] & $<.001$ \\
Mins since start & 0.98 & [0.98, 0.98] & $<.001$ \\
Mins since last tutoring & 1.02 & [1.02, 1.02] & $<.001$ \\
Is during tutoring & 1.61 & [1.56, 1.66] & $<.001$ \\
\midrule
\multicolumn{4}{l}{\textit{Random effects:}}\\
\multicolumn{4}{l}{$\sigma^2=0.31$;\quad $\tau$ (Student) $=0.18$;\quad \textit{ICC} $=0.37$}\\
\multicolumn{4}{l}{Marginal $R^2$ / Conditional $R^2 = 0.087 / 0.425$}\\
\bottomrule
\end{tabular}

\vspace{3pt}
\emph{Note.} \textit{IRR} $= e^{\beta}$; \textit{IRR}$>1$ indicates multiplicative increases in steps/min; 95\% \textit{CI}s are Wald. 
\end{table}

As shown in Table~\ref{tab:rq1_glmm}, students completed steps at a substantially higher rate during tutor visits (\textit{IRR} = 1.61, 95\% \textit{CI} [1.56, 1.66], $p<.001$). Engagement declined with session time (\textit{IRR} = 0.98, $p<.001$) but increased with time elapsed after a tutoring intervention (\textit{IRR} = 1.02, $p<.001$), leveling out the general decline. At reference values, the expected rate was 3.32 steps/min. Random-effects estimates (\textit{ICC} = .37) revealed considerable between-student heterogeneity; fixed effects explained 8.7\% of variance, rising to 42.5\% with random effects.

\subsection{Role of Intervention Length (RQ2)}

Across all tutoring episodes (N = 518), durations ranged from 1 to 43 minutes, with a median of 14 minutes and an interquartile range (IQR) of 14 minutes (Q1 = 7, Q3 = 21). The boxplot summary indicated a fairly symmetric distribution, with most episodes between 1 and 41 minutes. Only one episode extended beyond the conventional boxplot cutoff for high values (greater than 1.5 × IQR above Q3, here 42 minutes), and none fell below the cutoff for low values. Note that no episodes under 1 minute were included, as these very short sessions could not be linked to any MATHia log data and were thus excluded from the analysis.

Including standardized visit duration and its interaction with time since last tutoring significantly improved fit and parsimony relative to the visit-only model ($\Delta\mathrm{AIC}=1{,}279$; LR test $\chi^2(2)=1282.9$, $p<.001$). A one–SD longer visit was associated with a higher immediate engagement rate (\textit{IRR} $=1.07$, $95\%\,\mathrm{CI}=[1.07,\,1.08]$, $p<.001$). The interaction between time since last tutoring and duration was slightly below one (\textit{IRR} $=0.99$, $[0.98,\,0.99]$, $p<.001$), indicating a small acceleration of post-visit decay per additional minute since the visit for longer (vs.\ shorter) visits.

Other effects were robust and comparable to RQ1 (evaluated at mean duration): being during a visit remained positively associated with engagement (\textit{IRR} $=1.30$, $[1.25,\,1.35]$, $p<.001$), time since session start showed a small negative trend (\textit{IRR} $=0.98$, $[0.98,\,0.98]$, $p<.001$), and time since last tutoring retained a positive main effect (\textit{IRR} $=1.07$, $[1.06,\,1.07]$, $p<.001$).

\begin{figure*}[t]
  \centering
  \includegraphics[width=0.875\linewidth]{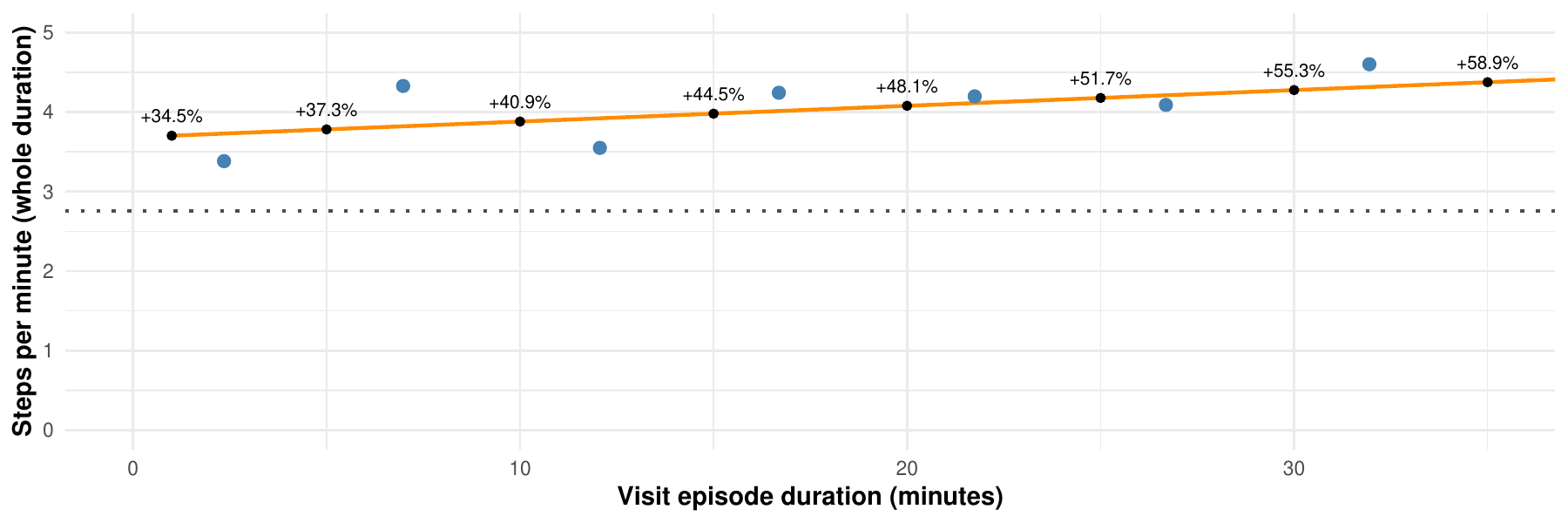}
  \caption{Engagement vs. tutoring episode duration. Blue points show empirical student-averaged steps/min for 5-minute duration bins; the orange line represents the estimated linear trend; the dotted line is the no-visit baseline. Labels indicate \% increase over baseline at example durations 1, 5, 10, 15, and so forth.}
  \label{fig:duration-curve}
\end{figure*}

\paragraph{Diminishing returns of longer visits.} 
To visualize these effects, we plotted the relationship between tutoring episode duration and engagement during the visit (Figure \ref{fig:duration-curve}). We first averaged steps per minute within student and binned visit durations into 5-minute intervals (0–5, 5–10, \dots, 35–40). We overlaid a linear regression trend line fit to episodes, and a dotted horizontal line marking the no-visit baseline (student-level mean). Black markers at 1, 5, 10, 15, 20, 25, 30, 35, and 40 minutes are annotated with the percent increase over the baseline.

The effect of longer tutoring episodes on engagement shows clear signs of diminishing returns: for example, a 10-minute visit was associated with a +41\% increase in average steps per minute over the entire visit duration, whereas a 20-minute visit yielded +48\%, meaning the second 10 minutes added only +7 percentage points on average; thus, while longer visits increase engagement more strongly throughout their duration and remain beneficial, the marginal gain per additional 10 minutes is smaller than the initial boost from starting a visit, suggesting that—when tutor time is limited—distributing shorter visits across students may be more efficient, though if tutors can stay, longer visits still maintain elevated engagement.

\paragraph{Post-visit trajectory (is the stronger decay substantive?).}
The interaction between episode duration and post-visit engagement decay was small and negative, meaning longer visits slightly \emph{speed up} the post-visit decay of the intervention. Although this effect slightly slows post-visit growth, model predictions indicate that it does not diminish the strong during-visit gains or the higher immediate post-visit level from longer visits. Specifically, after a typical visit, engagement rises by about $+6\%$ per minute; if the visit is $\sim$10 minutes longer, that rise is closer to $+3$--$4\%$ per minute, and if $\sim$20 minutes longer, it flattens to roughly $+1\%$ per minute but does not reverse the positive trend of visit duration. In other words, diminishing returns of visit duration (see previous paragraph) were stronger because students tended to show lower persistence after longer visits.

\subsection{Role of Intervention Timing (RQ3)}

\paragraph{Model evidence that later visits yield larger boosts (controlling for duration).}
To test whether the \emph{timing} of a tutor visit within a session matters—over and above visit length—we extended the RQ2 model to include interactions with minutes since session start. Adding these timing terms significantly improved fit and parsimony over the duration-only specification ($\Delta\mathrm{AIC}=96$; LR test $\chi^2(2)=100.8$, $p<.001$). The immediate during-visit effect increased later into the session (\textit{IRR}$_{\text{mins}\times\text{during}}=1.02$, $p<.001$), indicating that the same visit delivered later in a session is associated with a larger multiplicative lift in steps/min than an otherwise comparable visit delivered earlier. However, this growth was substantially smaller than the improvement per visit length (see previous model with an $IRR$ of 1.07). Consistent with this, descriptive averages across visit start times show progressively larger gains: the model predicts that visits beginning around 5 minutes show roughly a $+7.9\%$ increase compared to at the beginning of the session, whereas visits beginning around 30 minutes show about $+20.3\%$, and visits around 40 minutes approach $+25.2\%$.

By contrast, the post-visit growth rate showed a significantly negative slowing ($p<.001$) but the effect size was so small that it was virtually equivalent to no difference (\textit{IRR} $=1.00$; estimate $\approx-5.6\times10^{-4}$ on the log scale). This slowing is negligible relative to the during-visit increase.

\subsection{Difference Between Dialogue That Highly Increases Student Engagement vs. Not (RQ4)}

We first tested whether dialogue transcripts from high-uplift episodes differ systematically from lower-uplift ones (see Section~\ref{sec:rq4-randomization}). The observed centroid distance was $0.636$, significantly greater than 5{,}000 permutations, $p<.001$. A length-stratified variant confirmed the effect, $p<.001$. The classifier analysis likewise revealed a reliable signal: logistic regression with student-level cross-validation achieved an AUC of $0.69$, significantly above chance under permutation, $p<.001$. Together, these results show that high-uplift dialogues occupy a distinct embedding space, motivating qualitative analysis of their discourse features.

\paragraph{Thematic analysis: patterns in \emph{low-uplift} dialogues}

Guided by the quantitative screening (Section~\ref{sec:rq4-randomization}), we close-read exemplar episodes with the most negative model-based uplift and qualitatively coded recurrent interaction patterns tied to observable turns. Four low-impact categories emerged. First, some visits consisted almost entirely of unproductive talk or repeated technical checks (e.g., ``\emph{Can you hear me?}'') with little transition into task work. Second, tutors often offered brief affirmations or generic praise (``\emph{Excellent.}'') without following up with a question, prompt, or worked step, leaving students without direction. Third, episodes sometimes featured repeated mention of concepts (e.g., ``\emph{Cross multiplication.}'') without scaffolding or explanation, limiting opportunities for understanding. Finally, a small number of visits included silly or trivial remarks (e.g., ``\emph{Hello Mr. Cheese.}'') that neither acknowledged affect nor redirected to the task; although rare, these instances consistently coincided with below-baseline engagement.

\paragraph{Thematic analysis: patterns in \emph{high-uplift} dialogues} Exemplars from the highest model-based uplift revealed four groups. The most common involved concrete, stepwise scaffolding: tutors decomposed problems into executable moves (e.g., ``\emph{subtract 84 on both sides}'' $\rightarrow$ ``\emph{divide by 3}'' $\rightarrow$ ``\emph{state $x$}'') while naming relevant facts (``\emph{sum of angles is 180}'') and instantiating them with current numbers. Tutors also paired sensemaking checks (``\emph{Does that make sense?}'') with contingent next steps (``\emph{Now plug in the values you know}''), ensuring dialogue remained action-oriented. Another recurring pattern was making work visible: tutors organized the workspace, rewrote equations in canonical form, and labeled unknowns to reduce ambiguity. Finally, tutors often closed with actionable wrap-ups (e.g., ``\emph{Finish isolating $x$}'', ``\emph{Use the help button if stuck}''), preserving momentum beyond the visit.

\section{Discussion}

Our study is timely for learning analytics as hybrid human–AI classrooms increasingly rely on fine-grained traces to drive real-time orchestration. We examined how remote tutor visits align with ongoing work in an intelligent tutoring system and which dialogue features accompany stronger short-term gains in engagement. Three patterns stand out. First, visits coincide with reliable boosts in successful steps per minute, offsetting the general decline observed as sessions progress. Second, visit \emph{length} matters but with diminishing returns: longer episodes raise engagement during the visit, yet each added minute yields smaller incremental gains and persistence. Third, \emph{timing} matters: later check-ins generate larger immediate lifts than earlier ones, though an early pass remains useful to counter the downward drift of students who are never visited. Dialogue analyses further show that high-uplift episodes are marked by concrete, stepwise scaffolding and visible work organization, as opposed to generic praise without actionable next steps.

\subsection{What Do Brief Remote Visits Do?}

The sharp rise during a visit followed by a smaller, decaying carryover is consistent with human attention lowering barriers to progress in problem solving. Tutors disambiguate goals, clarify next steps, and externalize structure (e.g., rewriting an equation, labeling unknowns), which accords with classic accounts of scaffolding that reduce degrees of freedom and maintain task direction \cite{wood1976role}. A human–AI hybrid adaptivity lens further explains these patterns. AI tutors provide micro-adaptivity at scale—adjusting difficulty and delivering immediate feedback—yet often miss motivational and emotional factors that shape moment-to-moment engagement. Human tutors supply macro-adaptivity, guided by analytics that surface struggle or disengagement \cite{holstein2018student}. Spatiotemporal analyses likewise link human check-ins to improved in-system performance \cite{karumbaiah2023spatiotemporal}. Our results fit this picture: brief remote visits appear to lower the “activation energy” required to re-enter productive work within the AI environment.

That later visits show larger instantaneous lifts suggests time-varying friction: as fatigue and minor misconceptions accrue, clarifying structure becomes more valuable. A motivational account is also plausible: early in a session, attention and energy may be sufficient, whereas later, a human check-in functions as a salient reset that restores accountability and focus. Future work using affect detectors and field observations could adjudicate between these explanations \cite{miller2014boredom}.

\subsection{What \emph{Effective} Dialogue Looks Like}

The high- compared to low-uplift contrasts align with a significant body of literature on effective tutorial dialogue and instructional scaffolding. Our findings provide an empirical instantiation of expert tutoring principles identified in prior work. For instance, the "concrete, stepwise scaffolding" and "visible work organization" observed in high-uplift dialogues are direct parallels to the scaffolding moves detailed by Chi \cite{chi2001learning}, such as decomposing the task, highlighting critical features, and initiating a reasoning step. Similarly, our "sensemaking checks with contingent next steps" resonate with "Accountable Talk" frameworks, which emphasize moves like "Probing" and "Pressing for Reasoning" to get students to elaborate on their thinking \cite{abdelshiheed2024aligning,kupor2023measuring}.
These effective patterns stand in sharp contrast to the low-uplift dialogues, which were characterized by "phatic checks" or "praise without action." This distinction is consistent with frameworks like Wang and Demszky's \cite{wang2024tutor} "Tutor CoPilot," which empirically identified that less effective tutoring often involves generic encouragement or giving away the answer, rather than guiding the student's thinking process. The high-uplift moves we identified can be seen as the practical application of the broader principles for highly effective tutors described by Lepper and Woolverton \cite{lepper2002wisdom}, such as being socratic (eliciting information via questions), reflective (asking students to articulate their thinking), and progressive (moving from easier to more challenging cycles).

\subsection{Implications for Classroom Orchestration and Attention Allocation}

Our findings translate into actionable insights for resource-constrained hybrid and remote tutoring:

\begin{itemize}
    \item \textbf{Ensure at least one early touch.} A brief, early scan-and-redirect pass helps arrest the general decline in unvisited students and surfaces students who need follow-up.
    \item \textbf{Distribute time across students.} Because returns diminish with length, several short visits typically yield more total uplift than a few long ones, unless a student faces a deep impasse.
    \item \textbf{Target mid-to-late check-ins for larger lifts.} After an initial sweep, prioritize subsequent passes later in the session where the same minute of attention tends to buy more engagement.
    \item \textbf{Use “next-step scaffolds.”} End visits with a concrete, actionable waypoint (e.g., the next action in the tutoring system or subgoal) to extend momentum beyond the visit window.
\end{itemize}

These heuristics acknowledge the system-level trade-off—\emph{coverage} versus \emph{dosage}—and propose a policy that first stabilizes the room, then allocates later minutes where marginal returns are higher.

\subsection{Relation to Prior Work and Contribution to Learning Analytics}

Prior research shows that human attention and dialogue can raise engagement during tutoring. Classroom studies documented on-task gains when teachers attend to students \cite{person2003simulating} and spatiotemporal analytics traced how movement and noticing shape opportunities \cite{karumbaiah2023spatiotemporal}. Multimodal analytics has further demonstrated the value of integrating traces of interaction with learning outcomes \cite{d2015multimodal,blikstein2014multimodal}. Our results extend prior work in three respects. First, tutor visits produce immediate boosts in step completion rates that persist modestly afterward, even amid within-session decline. This minute-level alignment of dialogue and log data reveals engagement dynamics that would be obscured in aggregate analyses. Second, we disentangle the effects of visit \emph{length} and \emph{timing}, showing that longer visits yield diminishing returns while later visits deliver larger immediate lifts—implications for how scarce human attention might be scheduled across a session. Third, we link these quantitative patterns to dialogue content: high-uplift episodes feature concrete scaffolding and actionable steps, while low-uplift ones involve phatic checks or unguided praise.

These findings clarify how human support mitigates disengagement and suggest practical heuristics for classroom orchestration, such as distributing shorter visits, prioritizing mid-to-late check-ins, and ending with momentum-preserving prompts. More broadly, this work illustrates how learning analytics can move beyond prediction to explanatory accounts and actionable insights, providing a methodological template for integrating trace alignment, orchestration trade-offs, and dialogue analysis to support real-time instructional decision-making and equitable student support \cite{clow2012lac,siemens2013la}. In the long term, such approaches advance learning analytics by not only detecting and predicting learning behaviors but also explaining underlying mechanisms, guiding interventions, and promoting equitable outcomes. By linking fine-grained engagement dynamics to orchestration policies, this study advances a cumulative learning analytics science connecting micro-level traces to persistence, equity, and educational impact.

\subsection{Fairness Considerations for Orchestration}

Attention-allocation policies risk amplifying disparities if triage signals correlate with prior opportunity or visibility. The education community emphasizes ``humans in the loop'' for AI in education, keeping educators central \cite{ley2025teaching}. Orchestration tools should track unvisited students, audit uplift by demographics and achievement levels, and may enforce fairness constraints (e.g., minimum coverage) to avoid bypassing quieter students. Dialogue suggestions must be supportive rather than prescriptive to preserve tutor agency and cultural responsiveness.

\subsection{Limitations and Future Work}

Two tensions merit further study. First, the \emph{coverage vs. dosage} trade-off likely varies by class size, heterogeneity, and topic difficulty; micro-randomized trials (e.g., reinforcement learning-based A/B tests \cite{kizilcec2020scaling}) could yield context-adaptive schedules. Second, later visits' stronger immediate lifts may interact with end-of-class timing, since visits later into the session correspond to less time to unfold their effects. Future work should differentiate between opportunities for visits that are ``late but with runway'' and those that would be ``late and out of time.'' A key limitation is our joint treatment of student-tutor dialog: automatic diarization proved challenging, and manual checks showed $\geq 90\%$ tutor turns, so reported patterns likely emphasize tutor moves over student contributions. \revision{Associated future work may study students' reactions and perceptions of different tutor interventions from dialogue to better understand which tutor moves are most effective for motivation beyond observable engagement differences in log data.} Future analyses could isolate speakers and model their dynamics. More broadly, connecting momentary engagement gains to long-term \emph{learning} outcomes requires designs blending engagement-sensitive scheduling with skill-growth measures. \revision{We also acknowledge that our imperfect match rate of 73.3\% could have introduced selection bias. Finally, although participating teacher's in-class behavior is unlikely to confound our estimates of tutor effects---because tutor interactions typically occur while students are actively engaged with headphones on, making simultaneous teacher intervention improbable---it remains valuable to compare the \emph{tutor} effects observed here with the \emph{teacher} effects reported by past research \cite{karumbaiah2023spatiotemporal} in future work. Such a comparison can clarify whether in-the-moment support (from tutors versus teachers) relates to disengagement in similar ways and whether common principles of effective intervention emerge in technology-mediated tutoring settings (where, for instance, tutors can be more easily ignored by students than teachers in class).}

\section{Conclusion}

We present the first minute-level alignment of tutor–student dialogue with intelligent-tutor logs to examine \emph{how} brief human check-ins relate to momentary engagement during classroom practice. Visits coincide with immediate lifts in successful steps per minute and modest carryover. We disentangle \emph{length} from \emph{timing}: longer visits correspond to greater in-visit engagement but with diminishing returns, while later, well-timed passes are associated with larger instantaneous boosts even at comparable durations. We also find that concrete, stepwise scaffolding and explicit work organization are the dialogue features most consistently tied to higher productivity.

Integrating log traces with dialogue moves and content analysis reframes engagement as an actionable target for orchestration. Our methods pave the way for teacher- or tutor-facing analytics that recommend early stabilizing touch to curb decline, followed by short, strategically timed check-ins that sustain momentum under resource constraints. Content analytics could further recommend impactful dialogue moves while enforcing coverage constraints for equity.

Methodologically, we contribute a reproducible pipeline that aligns minute-level dialogue with system logs, models within-session dynamics without discarding the activity timeline, and links estimated uplift to dialogue content via embeddings and qualitative coding. The template generalizes to settings where human support complements AI tutors. By making the timing and form of human attention visible in traces, learning analytics can help allocate scarce support where it most effectively advances student learning through elevated levels of effort and engagement.

\begin{acks}
 This work was made possible with the support of the Learning Engineering Virtual Institute. The opinions, findings, and conclusions expressed in this material are those of the authors.
\end{acks}
        
\bibliographystyle{ACM-Reference-Format}
\bibliography{main}

\end{document}